\documentclass[%
 reprint,
superscriptaddress,
 amsmath,amssymb,
 aps,
]{revtex4-1}

\usepackage{graphicx}
\usepackage{dcolumn}
\usepackage{bm}
\usepackage{physics}
\usepackage{mathrsfs}

\begin{document}

\title{Quantum Discord Witness With Uncharacterized Devices}

\title{Quantum Discord Witness With Uncharacterized Devices}
\author{Rong Wang}
\email{ericrongwang19@gmail.com}
\affiliation{Department of Physics, University of Hong Kong, Pokfulam Road, Hong Kong SAR, China}
\author{Yao Yao}
\affiliation{Microsystems and Terahertz Research Center, China Academy of Engineering Physics, Chengdu Sichuan 610200, China}
\author{Zhen-Qiang Yin}
\email{yinzq@ustc.edu.cn}
\affiliation{CAS Key Laboratory of Quantum Information, University of Science and Technology of China, Hefei 230026, China}
\affiliation{CAS Center for Excellence in Quantum Information and Quantum Physics, University of Science and Technology of China, Hefei 230026, China}
\affiliation{Hefei National Laboratory, University of Science and Technology of China, Hefei 230088, China}





\begin{abstract}
Quantum discord represents a correlation beyond classicality that could be useful for many quantum information tasks, and therefore is viewed as a valuable quantum resource. Theoretically, whether a bipartite quantum state has a non-zero amount of quantum discord or not can be determined once its density matrix is given. Experimentally, however, it is not easy to perform a reliable tomography on the state, especially when considering the imperfection of devices and the high dimensionality of system. Here, inspired by the so-called dimension witness, we propose a new approach using uncharacterized measurements to witness quantum discord of an unknown bipartite state within arbitrary dimension system. Furthermore, $2 \times 2$ measurements are sufficient for any bipartite state of arbitrarily large dimension. For any two-qubit state, we show that the maximum of witness happens when one performs mutually orthogonal basis measurement on each qubit. Our method exhibits high robustness against device imperfections, such as error tolerance, indicating its experimental feasibility.
\end{abstract}

\pacs{Valid PACS appear here}
\maketitle


\onecolumngrid

\section{\label{sec:level1}Introduction}

\noindent 

Quantum discord (QD) is one type of correlation of quantumness \cite{ollivier2001quantum,modi2012classical}, and unlike entanglement \cite{horodecki2009quantum}, exists in a wide rang of bipartite quantum states. Since QD shows its power in diversified tasks, such as mixed state quantum computing \cite{knill1998power,datta2008quantum}, remote state preparation \cite{dakic2012quantum}, coherent quantum interactions \cite{gu2012observing}, bounding distributed entanglement \cite{chuan2012quantum}, quantum channel discrimination \cite{weedbrook2016discord}, quantum metrology \cite{girolami2013characterizing}, quantum state merging \cite{cavalcanti2011operational,madhok2011interpreting}, and other quantum information processes \cite{modi2012classical}, researchers have paid much attention to it in last two decades. One attractive characteristic of QD is that it is not as fragile as entanglement. For example, it has been proven that, for almost all positive-discord states, discord cannot vanish suddenly and permanently at a finite time \cite{yu2009sudden}, which is totally different from entanglement \cite{ferraro2010almost}. Therefore, it can to some extent be preserved in a noisy environment. One other characteristic is that, even a separable state may carry a non-zero amount of QD and thus, can be useful for achieving various quantum information tasks. 

Like other quantum resources, it is of significant importance to detect QD. State tomography always works out, but it requires sufficient and accurate measurements. For instance, the number of measurement grows squarely when it comes to a $d \times d$ dimensional bipartite systems. 
The demand of practical implementations sometimes limits the power of state tomography since real-world devices would not always work well, and therefore a much more practical approach to detect or witness the QD is in need.

Since QD is a kind of non-convex resource \cite{chitambar2019quantum,bera2017quantum}, a linear witness operator cannot detect it successfully when it comes to a separable state \cite{rahimi2010single}. Thus, a good witness should be nonlinear. Researchers have proposed several approaches \cite{bylicka2010witnessing,gessner2013local,zhang2011detecting,maziero2012classicality,girolami2012observable} towards the QD witness and experimentally demonstrated it \cite{xu2010experimental,passante2011experimental,aguilar2012experimental,silva2013measuring}. However, all these approaches require perfect measurements. Up till now, a method to witness QD with uncharacterized devices is still missing. 

Here, we solve this problem directly by proposing a method to witness quantum discord with uncharacterized devices. Inspired by dimension witness in a prepare-and-measure scenario with independent devices \cite{bowles2014certifying}, we propose a nonlinear witness that only requires three natural assumptions: i) source and measurement devices are independent, ii) the dimension of the system is given, iii) the state and measurements are independently and identically distributed (i.i.d) in each experimental round. In this paper, we first discuss the simplest case of two-qubit discord witness, and then generalize it to arbitrary dimension. In the case of two-qubit, we analyze a model of Bell-like experiment (see Fig. \ref{fig:setup}), and present the sufficient condition of successful witness. Furthermore, given any density matrix of two qubits, we construct explicitly the measurement operators that enable the discord witness to achieve a maximal value. 

In this work, we also show that our approach is robust, that is, the witness value decrease linearly with the errors of prepared states and measurements. More importantly, the required number of measurements will always be $2 \times 2$ for any bipartite system, which is independent of the dimension. These features shows our demands on devices are quite loose and our approach is extremely robust to device imperfections.

\section{Model} 

\begin{figure}[htbp]
\centering
\includegraphics[width=0.5\linewidth]{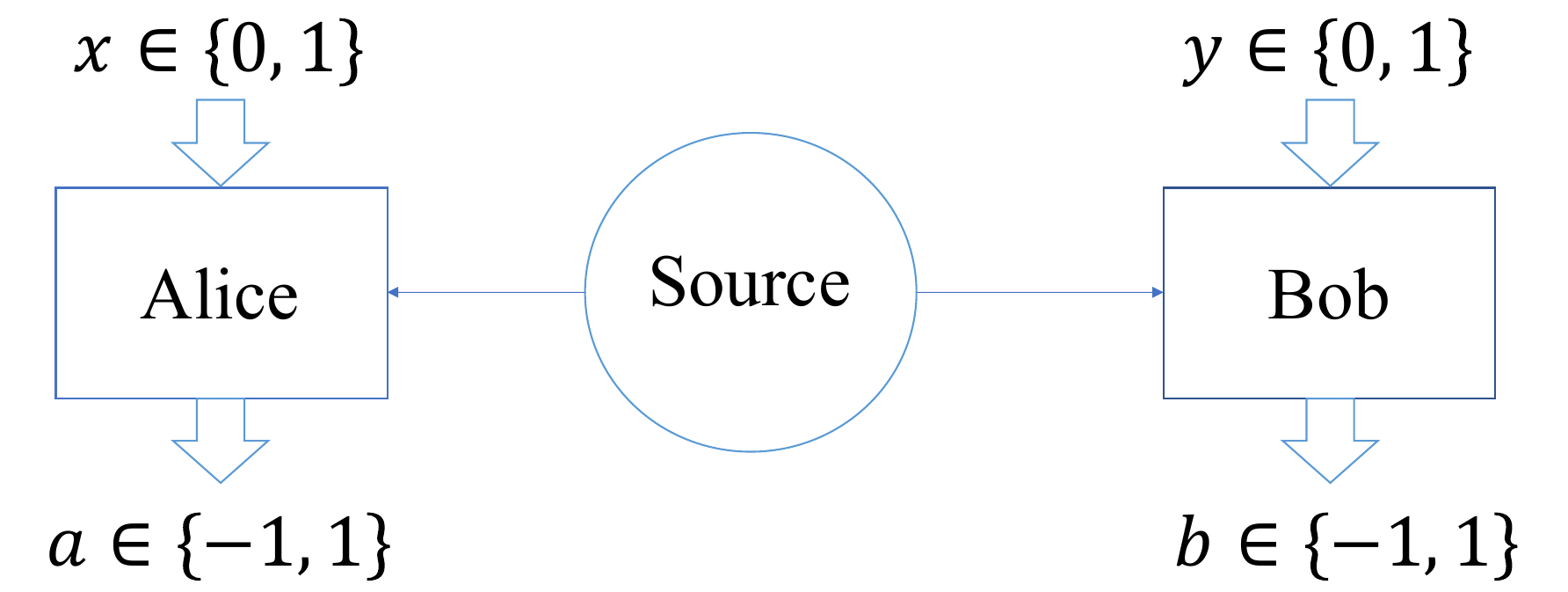}
\caption{Quantum discord witness set-up.}
\label{fig:setup}
\end{figure}

We consider the two-qubit witness experiment described by following steps. Step 1: in each round, a source prepares an unknown two-qubit state $\rho_{AB}$, and sends each of them to Alice and Bob respectively. Step 2: Alice (Bob) randomly chooses $x \in \{0, 1\}$ ($y \in \{0, 1\}$), performs a random two-dimensional measurement $A_x$ ($B_y$) on her (his) qubit and obtains the outcome $a \in \{-1, 1\}$ ($b \in \{-1, 1\}$). Step 3: after all rounds have finished, Alice and Bob collect the data and calculate the expectation values $\left \langle A_x \right \rangle = \Tr[\rho_{A}A_x ]$, $\left \langle B_y \right \rangle =\Tr[\rho_{B}B_x ]$ and $\left \langle A_x \otimes B_y \right \rangle = p(a = b|xy)-p(a \ne b|xy)=\Tr[\rho_{AB}A_x \otimes B_y]$, where $\rho_A$ ($\rho_B$) is Alice's (Bob's) subsystem. The brief set-up is shown in Fig. \ref{fig:setup}. In the case of the two-qubit discord witness, the three assumptions in our approach have been introduced before, with that the second assumption being modified to, ii') the subsystems of Alice and Bob are two-dimensional. The second and third assumptions are easy to understand, here we briefly review the first assumption \cite{bowles2014certifying}: considering a hidden variable model where the observables $A_x (\lambda_a)$, $B_y (\lambda_b)$ and the state $\rho_{AB}(\lambda_s)$ are respectively controlled by hidden variables $\lambda_a$, $\lambda_b$ and $\lambda_s$, respectively, and these hidden variables are unknown to the observer. They satisfy $\int d \lambda_a q(\lambda_a)=\int d \lambda_b r(\lambda_b)=\int d \lambda_s t(\lambda_s)=1$ where $q(\lambda_a)$, $r(\lambda_b)$ and $t(\lambda_s)$ are marginal probability density functions, and $A_x=\int d \lambda_a q(\lambda_a)A_x (\lambda_a)$, $B_y=\int d \lambda_b r(\lambda_b)B_y (\lambda_b)$ and $\rho_{AB}=\int d \lambda_s t(\lambda_s) \rho_{AB}(\lambda_s)$, respectively. Then the independence of all devices means that the global probability density function denoted by $p(\lambda_a \lambda_b \lambda_s)=q(\lambda_a)r(\lambda_b)t(\lambda_s)$, thus the observable probability is given by 
\begin{equation}
\begin{aligned}
   &\int d \lambda_a d \lambda_b d \lambda_s  p(\lambda_a \lambda_b \lambda_s) \Tr[\rho_{AB}(\lambda_s) A_x(\lambda_a) \otimes B_y(\lambda_b)]  \\
 =& \int d \lambda_a d \lambda_b d \lambda_s q(\lambda_a)r(\lambda_b)t(\lambda_s) \Tr[\rho_{AB}(\lambda_s) A_x(\lambda_a) \otimes B_y(\lambda_b)] \\
 =& \Tr[\rho_{AB}A_x \otimes B_y], \\
\end{aligned}
\end{equation}
and we have same arguments on $\left \langle A_x \right \rangle$ and $\left \langle B_y \right \rangle$. 

Furthermore, let’s explain the reason why hidden variable model should be considered. Since our discord witness is based on the uncharacterized devices, it must have a chance to argue that the state and the measurements have pre-shard resources. That is to say the state and the measurements may correlate with each other before the witness. In this case, our witness fails. Here, we take a simple example to demonstrate it. Let $\lambda \in \{0,1\}$ be the global hidden variable with uniformly distribution. If $\lambda=0$, the source prepares the Bell state $\ket{\phi^+}:=(\ket{00}+\ket{11})/\sqrt{2}$, Alice’s (Bob’s) measurements are Pauli Z-measurement $A_0(B_0)=\mathbb{Z}$ and Pauli X-measurement $A_1(B_1)=\mathbb{X}$. If $\lambda=1$, the source prepares another Bell state $\ket{\phi^-}:=(\ket{00}-\ket{11})/\sqrt{2}$, Alice’s measurements are still $A_0=\mathbb{Z}$ and $A_1=\mathbb{X}$, while Bob’s measurements are $B_0=\mathbb{Z}$ and $B_1=-\mathbb{X}$. By a simple calculation, witness value defined in Eq. \eqref{W_value} is $1$, indicating a non-zero discord state. However, if looking at the state only, it is a uniform mixture of $\ket{\phi^+}$ and $\ket{\phi^-}$, resulting in $\rho_{AB}=(\ket{00}\bra{00}+\ket{11}\bra{11})/2$, which is obviously a zero-discord state. This example demonstrates that the state and the measurements must have no correlations before witness, otherwise, our method finals. In previous works, such as Refs. \cite{bylicka2010witnessing,gessner2013local,zhang2011detecting,maziero2012classicality,girolami2012observable}, their methods are all based on perfect and characterized measurements, so that it’s no need to take hidden variable model into account. In the contrast, our method doesn’t require characterized devices, which allows the existence of such correlation. Hence, we need to assume the independence of the involved devices. This fact also reflects the non-triviality of our method.



\section{Main result}

Here, we first present the witness form and then derive that any zero-discord two-qubit state always leads to zero witness value. We define $Q_{xy}:= \left \langle A_x \otimes B_y \right \rangle - \left \langle A_x \right \rangle \left \langle B_y \right \rangle $, then the witness value is given by the following determinant
\begin{equation}
\label{W_value}
\begin{aligned}
W:=
\begin{vmatrix}
Q_{00} & Q_{01}  \\
Q_{10} & Q_{11}
\end{vmatrix}.
\end{aligned}
\end{equation}
In Bloch representation, we have 
\begin{equation}
\rho_{AB}=\frac{1}{4}(\mathbb{I} \otimes \mathbb{I} +\vec{S}_a \cdot \vec{\sigma} \otimes \mathbb{I} + \mathbb{I} \otimes \vec{S}_b \cdot \vec{\sigma} + \vec{\sigma}^t \hat{T} \vec{\sigma}),
\end{equation}
where $\mathbb{I}$ is the identity matrix, $\vec{\sigma}=(\sigma_x, \sigma_y, \sigma_z)^t $ is the vector of Pauli matrix, $\vec{S}_a$ and $\vec{S}_b$ are the local Bloch vectors for $\rho_A=\frac{1}{2}(\mathbb{I}+\vec{S}_a \cdot \vec{\sigma}$) and $\rho_B=\frac{1}{2}(\mathbb{I}+\vec{S}_b \cdot \vec{\sigma}$), respectively, and $\hat{T}$ is the $3 \times 3$ correlation tensor matrix. For measurement operator, $A_x=a_x\mathbb{I}+\vec{A}_x \cdot \vec{\sigma}$ and $B_y=b_y\mathbb{I}+\vec{B}_y \cdot \vec{\sigma}$, where $a_x$ and $b_y$ are constants, $\vec{A}_x$ and $\vec{B}_y$ are also Bloch vectors with constraints $|a_x| \le 1-|\vec{A}_x|$ and $|b_x| \le 1-|\vec{B}_y|$. Then we have 
\begin{equation}
\left \langle A_x \otimes B_y \right \rangle=a_x b_y +\vec{S}_a \cdot \vec{A}_x b_y + a_x \vec{S}_b \cdot \vec{B}_y + \vec{A}^t_x \hat{T} \vec{B}_y.
\end{equation}
Similarly, 
\begin{equation}
\left \langle A_x \right \rangle \left \langle B_y \right \rangle = (a_x+\vec{S}_a \cdot \vec{A}_x) (b_y + \vec{S}_b \cdot \vec{B}_y),
\end{equation}
then, we have
\begin{equation}
Q_{xy} = \vec{A}^t_x \hat{T} \vec{B}_y - (\vec{S}_a \cdot \vec{A}_x) (\vec{S}_b \cdot \vec{B}_y)=\vec{A}^t_x (\hat{T}-\vec{S}_a\vec{S}^t_b) \vec{B}_y .
\end{equation}
To simplify, we define a $3 \times 3$ matrix $\hat{S}:=\hat{T}-\vec{S}_a\vec{S}^t_b$, as a consequence, we have
\begin{equation}
\label{W}
\begin{aligned}
W&= 
\begin{vmatrix}
\vec{A}^t_0 \hat{S} \vec{B}_0 & \vec{A}^t_0 \hat{S} \vec{B}_1  \\
\vec{A}^t_1 \hat{S} \vec{B}_0 & \vec{A}^t_1 \hat{S} \vec{B}_1
\end{vmatrix} 
=(\vec{A}_0 \times \vec{A}_1) \cdot (\hat{S}\vec{B}_0 \times \hat{S}\vec{B}_1). \\
\end{aligned}
\end{equation}
It is straightforward to see that the witness would vanish if the rank of $\hat{S}$ is one or zero. Then we shall show that any zero-discord two-qubit state will always be in that case. As proved in \cite{datta2008studies}, the zero-discord state can be written as $\chi_{AB}=p_0\ket{0}\bra{0} \otimes \rho_0+p_1\ket{1}\bra{1} \otimes \rho_1$, where
$\{\ket{0}, \ket{1}\}$ is  an orthonormal basis of Alice's system, up to a local unitary transformation on Alice's side, $\rho_{0, 1}$ are $2 \times 2$ density matrices of Bob's system. In Bloch representation, it is given by
\begin{equation}
\begin{aligned}
\chi_{AB} =& \frac{1}{4}(\mathbb{I} \otimes \mathbb{I} + (p_0-p_1) \sigma_z \otimes \mathbb{I} 
                  +\mathbb{I} \otimes (p_0\vec{s}_0+p_1\vec{s}_1) \cdot \vec{\sigma} 
                 +\sigma_z \otimes (p_0\vec{s}_0-p_1\vec{s}_1) \cdot \vec{\sigma}), \\
\end{aligned}
\end{equation}
where $\vec{s}_{0 ,1}$ are the Bloch vectors for $\rho_{0, 1}=\frac{1}{2}(\mathbb{I}+\vec{s}_{0, 1} \cdot \vec{\sigma})$. After simple calculation, the matrix $\hat{S}$ for $\chi_{AB}$ is given by
\begin{equation}
\hat{S}_{\chi} := (\vec{0}, \vec{0}, 2p_0p_1(\vec{s}_0-\vec{s}_1))^t, 
\end{equation}
where the three elements of both first row and second row are all zero, and thus the rank of $\hat{S}_{\chi}$ is one or zero. Note that the rank of $\hat{S}$ is invariant under local unitary on Alice's side. Besides, the construction of this witness is symmetric, thus, it can’t be non-zero as long as the discord on either Alice or Bob is zero. So, we conclude that a non-zero value of $W$ is a sufficient condition for non-zero discord, therefore, the zero-discord two-qubit states always have a vanishing witness.

In the following, given any density matrix of two qubits, we find explicitly the measurement operators that allow the discord witness to achieve a maximal value, provided a fixed $\rho_{AB}$. Let $\vec{i}=\vec{B}_0/|\vec{B}_0|$ and $\vec{j}=\vec{B}_1/|\vec{B}_1|$ be the unit vectors respectively, $k_1$ and $k_2$ are the first and second largest eigenvalues of matrix $\sqrt{\hat{S}\hat{S}^t}$ respectively, using the fact that Bloch vectors $0 \le |\vec{A}_x| \le 1$ and $0 \le |\vec{B}_y| \le 1$, we have
\begin{equation}
\begin{aligned}
W & \le |\vec{A}_0 \times \vec{A}_1| \cdot |\hat{S}\vec{B}_0 \times \hat{S}\vec{B}_1| \\
    & \le |\hat{S}\vec{B}_0 \times \hat{S}\vec{B}_1| \\
    & \le |\hat{S}\vec{i} \times \hat{S}\vec{j}| \\
    & \le k_1k_2, \\
\end{aligned}
\end{equation}
where the equality of the first inequality holds if vectors $\vec{A}_0 \times \vec{A}_1$ and $\hat{S}\vec{B}_0 \times \hat{S}\vec{B}_1$ are in the same direction, the equality of the second inequality holds if $\vec{A}_x$ are mutually orthogonal with $|\vec{A}_x|=1$, the equality of the third inequality holds if $|\vec{B}_y|=1$, and the equality of the last inequality holds if $\vec{B}_y$ are mutually orthogonal and the plane formed by eigenvectors of $k_1$ and $k_2$ is the same as the plane formed by $\vec{B}_y$. For the last inequality in above formula, we have made use of the fact that
\begin{equation}
\begin{aligned}
|\hat{S}\vec{i} \times \hat{S}\vec{j}| & = \frac{1}{2}|\hat{S} (\vec{i}+\vec{j}) \times \hat{S}(\vec{i}-\vec{j})| \\ 
                                                        & = \left| 2\cos\theta \sin\theta \frac{\hat{S} (\vec{i}+\vec{j})}{|\vec{i}+\vec{j}|} \times \frac{\hat{S} (\vec{i}-\vec{j})}{|\vec{i}-\vec{j}|} \right| \\
                                                        & \le \left| \frac{\hat{S} (\vec{i}+\vec{j})}{|\vec{i}+\vec{j}|} \times \frac{\hat{S} (\vec{i}-\vec{j})}{|\vec{i}-\vec{j}|} \right|, \\
\end{aligned}
\end{equation}
where $2\theta \in (-\pi/2, \pi/2]$ is the angle of vectors $\vec{i}$ and $\vec{j}$, we make use of the inequality $2\cos\theta \sin\theta \le \cos^2\theta + \sin^2\theta=1$ and thus the equality holds if $\theta=\pi/4$, that is, $\vec{i}$ and $\vec{j}$ are mutually orthogonal. Thus, witness value comes to maximum if and only if $\vec{A}_x$ as well as $\vec{B}_y$ are some apposite mutually orthogonal unit vectors, depending on matrix $\hat{S}$. In other word, the measurements should be two mutually orthogonal basis (MUB) on both Alice's and Bob's sides. 

\section{High-dimensional case}

In this section, we generalized the main result to high-dimensional case. Here we consider the zero-discord state $\chi_{AB}=\sum_{i=0}^{d_A-1} p_i\ket{i}\bra{i} \otimes \rho_i$ and the similar experiment scenario with inputs $x, y \in \{0, 1, \cdots, d_A-1\}$ and outputs $a, b \in \{-1, 1\}$, where $d_A$ is the dimension of Alice's system. Similarly, we construct the $d_A \times d_A$ matrix $\hat{W}_{d_A}$ and its element at row $x$, column $y$ that $Q_{xy}= \left \langle A_x \otimes B_y \right \rangle - \left \langle A_x \right \rangle \left \langle B_y \right \rangle $. Similarly, the witness value is given by the determinant $W_{d_A}=\det(\hat{W}_{d_A})$. Then we choose basis sets of local Hermitian operators $\{ \tau_1, \cdots, \tau_i, \cdots, \tau_{d^2_{A}-1} \}$ ($\{ \upsilon_1, \cdots, \upsilon_j, \cdots, \upsilon_{d^2_{B}-1} \}$), satisfying $\Tr[\tau_i \tau_{i^{'}}]=2\delta_{i i^{'}}$ ($\Tr[\upsilon_j \upsilon_{j^{'}}]=2\delta_{j j^{'}}$). Then any bipartite state is given by
\begin{equation}
\begin{aligned}
\rho_{AB} =& \frac{1}{d_A d_B}( \mathbb{I}_{d_A} \otimes \mathbb{I}_{d_B} + \sqrt{\frac{d_A(d_A-1)}{2}} \vec{S}_a \cdot \vec{\tau} \otimes \mathbb{I}_{d_B}  \\
&+ \sqrt{\frac{d_B(d_B-1)}{2}} \mathbb{I}_{d_A} \otimes \vec{S}_b \cdot \vec{\upsilon} + \sqrt{\frac{d_A d_B(d_A-1)(d_B-1)}{4}} \vec{\tau}^t \hat{T} \vec{\upsilon}), \\
\end{aligned}
\end{equation}
where $\vec{S}_a \in \mathbb{R}^{d^2_A-1}$ ($\vec{S}_b \in \mathbb{R}^{d^2_B-1}$), $|\vec{S}_a| \le 1$ ($|\vec{S}_b| \le 1$), $\vec{\tau}=(\tau_1, \cdots, \tau_{d^2_A-1})^t$ ($\vec{\upsilon}=(\upsilon_1, \cdots, \upsilon_{d^2_B-1})^t$) is the vector of the $d^2_A-1$ ($d^2_B-1$) generalized Gell-Mann matrices. For measurement operator, we have $A_x=\frac{2}{d_A }(a_x\mathbb{I}_{d_A}+  \sqrt{\frac{d_A(d_A-1)}{2}} \vec{A}_x \cdot \vec{\tau})$ ( $B_y=\frac{2}{d_B}(b_y\mathbb{I}_{d_B}+  \sqrt{\frac{d_B(d_B-1)}{2}} \vec{B}_y \cdot \vec{\upsilon})$), where $\vec{A}_x \in \mathbb{R}^{d^2_A-1}$ ($\vec{B}_y \in \mathbb{R}^{d^2_B-1}$) and $|\vec{A}_x| \le 1$ ($|\vec{B}_y| \le 1$). Then we obtain the element 
\begin{equation}
Q_{xy} =\frac{4(d_A-1)(d_B-1)}{d_Ad_B} \vec{A}^t_x (\hat{T}-\vec{S}^t_a\vec{S}_b) \vec{B}_y .
\end{equation}
Thus, as above, we get the $(d^2_A-1) \times (d^2_B-1)$ matrix $\hat{S}=\hat{T}-\vec{S}_a\vec{S}^t_b$, and similarly to Eq. \eqref{W}, the generalized witness of arbitrary dimensions can be expressed by using cross product that 
\begin{equation}
\begin{aligned}
W_{d_A}& =\det(\hat{W}_{d_A}) = \left[\frac{4(d_A-1)(d_B-1)}{d_Ad_B} \right]^{d_A} \vec{A} \cdot \vec{B} , \\
\end{aligned}
\end{equation}
where $\vec{A}:=\vec{A}_0 \times \vec{A}_1 \times\cdots \times \vec{A}_{d_A-1}$ and $\vec{B}:=\hat{S}\vec{B}_0 \times \hat{S}\vec{B}_1 \times\cdots \times \hat{S}\vec{B}_{d_A-1}$
of $d_A$ vectors are defined in $\mathbb{R}^{d_A+1}$ \cite{bowles2014certifying,dittmer1994cross}. Therefore, we immediately have that $\vec{B}=\vec{0}$ if and only if $\{\hat{S}\vec{B}_0, \cdots, \hat{S}\vec{B}_{d_A-1}\}$ are linearly dependent, and this condition can be guaranteed if the rank of $\hat{S}$ is not greater than $d_A-1$. Then we shall prove that the rank of the matrix $\hat{S}$ for $\chi_{AB}$  is bounded by $d_A-1$. In the representation of Hermitian operator, we define $\ket{i}\bra{i}=\frac{1}{d_A}(\mathbb{I}_{d_A} + \sqrt{\frac{d_A(d_A-1)}{2}} \vec{r}_i \cdot \vec{\tau})$ with $1+ (d_A-1)\vec{r}_i \cdot \vec{r}_{i^{'}}=0$ (if $i \neq i^{'}$) and $\rho_i=\frac{1}{d_B}(\mathbb{I}_{d_B}+ \sqrt{\frac{d_B(d_B-1)}{2}} \vec{s}_i \cdot \vec{\upsilon}_i)$, then we have
\begin{equation}
\begin{aligned}
\chi_{AB} =& \frac{1}{d_A d_B}( \mathbb{I}_{d_A} \otimes \mathbb{I}_{d_B}  + \sqrt{\frac{d_A(d_A-1)}{2}} \sum_i p_i \vec{r}_i \cdot \vec{\tau} \otimes \mathbb{I}_{d_B} \\
&+ \sqrt{\frac{d_B(d_B-1)}{2}} \mathbb{I}_{d_A} \otimes \sum_i p_i \vec{s}_i \cdot \vec{\upsilon} +\sqrt{\frac{d_A d_B(d_A-1)(d_B-1)}{4}}\vec{\tau}^t \sum_i p_i \vec{r}_i \vec{s}^t_i \vec{\upsilon}). \\
\end{aligned}
\end{equation}
As above, we have the matrix $\hat{S}$ for $\chi_{AB}$ given by
\begin{equation}
\begin{aligned}
\hat{S}_{\chi} =& \sum_i p_i \vec{r}_i \vec{s}^t_i- (\sum_i p_i \vec{r}_i) (\sum_{i^{'}} p_{i^{'}} \vec{s}^t_{i^{'}}) \\
                      =& \sum_i p_i \vec{r}_i (\vec{s}_i- \sum_{i^{'}} p_{i^{'}} \vec{s}_{i^{'}})^t, \\
\end{aligned}
\end{equation}
where we can see that all the row vectors contained in the vector space spanned by $\vec{s}_i- \sum_{i^{'}} p_{i^{'}} \vec{s}_{i^{'}}$ for all $i \in \{0, \cdots, d_A-1\}$. It is straightforward to see that  these vectors are linearly dependent since we can make use of the probabilities $\{p_i\}$ with $\sum_i p_i=1$, resulting in 
\begin{equation}
\begin{aligned}
\sum_i p_i \vec{s}_i- \sum_{i^{'}} p_{i^{'}} \vec{s}_{i^{'}}=0. 
\end{aligned}
\end{equation}
Therefore, the dimension of the vector space spanned by $\vec{s}_i- \sum_{i^{'}} p_{i^{'}} \vec{s}_{i^{'}}$ is at most $d_A-1$, that is, the rank of $\hat{S}_{\chi}$ is bounded by $d_A-1$, which completes the proof. The key point of this our method is to make use of the fact that the rank of matrix $\hat{S}$ is always bounded by $d_A-1$ for every zero-discord state, and thus, for any $d_A-1$ linearly independent vectors $ \{\vec{B}_0, \cdots, \vec{B}_{d_A-1} \}$, $\{\hat{S}\vec{B}_0, \cdots, \hat{S}\vec{B}_{d_A-1}\}$ are always linearly dependent. We conclude that the zero-discord states of arbitrary dimension always have a vanishing witness.

\section{Measurement with multiple outcomes} 

In the previous section, we assume that any measurement yields binary outcomes in high-dimensional cases. For a $d \times d$ bipartite system, at least $d$ measurements are required on each side, making this approach as costly as the tomographic method. However, the situation changes when we consider measurements with multiple outcomes. Initially, it was believed that high-dimensional measurements with multiple outcomes could be implemented practically, such as in photon systems \cite{erhard2020advances}. Next, we consider measurements with $d$ outcomes, a typical example is the projective measurement for $d$-dimensional state.  Let $\{M_i\}$ with $i \in \{0, 1, \cdots, d-1\}$ be the elements of POVMs of a $d$-outcome measurement, and $M_i=\frac{1}{d} (m_i\mathbb{I}_{d}+  \sqrt{\frac{d(d-1)}{2}} \vec{M}_i \cdot \vec{\mu})$ \cite{bowles2014certifying}, where $m_i$ is a real number, $\vec{M}_i \in \mathbb{R}^{d^2-1}$ and $\vec{\mu}=(\mu_1, \cdots, \mu_{d^2-1} )^t$ is the vector of Hermitian operators. We assume $\{M_i\}$ are good enough, that is, the any $d-1$ vectors of $\{ \vec{M}_i \}$ are linearly independent. Then we could exploit $\{\vec{M}_i\}$ to construct at most $d-1$ linearly independent vectors, for example, 
\begin{equation}
\vec{N}_j=\sum_{i=0}^{j} \vec{M}_i - \sum_{i=j+1}^{d-1} \vec{M}_i,
\end{equation}
with $j \in \{1, \cdots, d-1\}$. Correspondingly, the measurement operators are given by $N_j=\sum_{i=0}^{j} M_i - \sum_{i=j+1}^{d-1} M_i$. We can derive $d-1$ binary-outcome measurements from a single $d$-outcome measurement. This concept is based on the fact that quantum measurements can be manipulated classically through post-processing \cite{buscemi2005clean,d2005classical,haapasalo2012quantum,oszmaniec2017simulating}. In this example, we retag the outcomes $\{0, \cdots, j\}$ as $1$ and the outcomes $\{j+1, \cdots, d-1\}$ as $-1$, resulting in a new POVM for each $j$. By this method, we have already generated $d-1$ binary-outcome measurements. According to our theory, we only need one additional binary-outcome measurement, which must have a corresponding vector that is linearly independent of $\{\vec{N}_j\}$. This binary-outcome measurement can also be obtained through a multiple-outcome measurement followed by retagging. In this way, we actually require two measurements: the $d$-outcome measurement and an additional binary-outcome measurement.

To clearly demonstrate the effectiveness of this method, we consider a two-pair qubit system as a high-dimensional example to evaluate the witness value. Let $\rho_{AB}^2=p\ket{\phi^+}\bra{\phi^+}^{\otimes 2}+\frac{1-p}{8}\mathbb{I}_4^{\otimes 2}$ be high-dimensional Werner state. The chosen $4$-outcome measurement is the projective measurement using the computational basis $\{\ket{00}, \ket{01},  \ket{10}, \ket{11}\}$. On Alice's side, an alternative way to construct $3$ binary-outcome measurements operators is as follows
\begin{equation}
\begin{aligned}
A_1& =\ket{00}\bra{00}+\ket{01}\bra{01}-\ket{10}\bra{10}-\ket{11}\bra{11} \\
A_2& =\ket{00}\bra{00}-\ket{01}\bra{01}+\ket{10}\bra{10}-\ket{11}\bra{11} \\
A_3& =\ket{00}\bra{00}-\ket{01}\bra{01}-\ket{10}\bra{10}+\ket{11}\bra{11}. \\
\end{aligned}
\end{equation}
The chosen binary-outcome measurement is given by 
\begin{equation}
A_0 =(\ket{0}\bra{1}+\ket{1}\bra{0}) \otimes \mathbb{I}_2.
\end{equation}
The measurements are analogous on Bob's side. By standard calculation, $Q_{xy}=p$ if $x=y$ ($x, y \in \{0,1, 2, 3\}$ in this exmaple), otherwise, $Q_{xy}=0$. Consequently, the corresponding witness value is $W_{\text{two-pair}}=p^4$. As long as $p$ is positive, $W_{\text{two-pair}}$ will not vanish.

Furthermore, we consider the case of $n$-pair of qubit represented by $\rho_{AB}^n$ with $n \in \mathbb{Z}^+$. As indicated in the previous example, we could perform local qubit measurement to construct a specific projective measurement. For example, Alice (Bob) perform local Pauli $\mathbb{Z}$ measurement on her (his) all $n$ qubits. An alternative binary-outcome measurement could be a local Pauli $\mathbb{X}$ measurement on just one qubit. This approach makes it practical to witness the discord of $\rho_{AB}^n$, as we only need local qubit measurements on both Alice's and Bob's sides. In terms of the number of local qubit measurements, $n+1$ measurements on each side are sufficient. This means that the local measurement scales linearly, while the dimension of $d=2^n \times 2^n$ grows exponentially.

\section{Robustness analysis}

In this section, we use the two-qubit witness as an example to analyze the robustness of our method, which can be naturally extended to the high-dimensional case. In our context, robustness refers to the witness value changing slowly in response to variations in state preparation and measurements. Furthermore, we will demonstrate that the deviation of the witness is a linear function of the deviations in the inputs, specifically the state preparation and the measurements. This also indicates the continuity of our discord witness.

Let's define a family $\Phi[\rho_{AB}, A_0, A_1, B_0, B_1]$ of two-qubit witness experiment, then the witness value $W$ could be viewed as the function of $\Phi$, denoted as $W(\Phi)$. In practical implementation, each element in $\Phi$ could occur error, which results from the background noise. We define a family $\Phi'[\rho'_{AB}, A'_0, A'_1, B'_0, B'_1]$ deviated from $\Phi$. The the robustness analysis aims at quantifying the upper bound of difference $\delta:=|W(\Phi')-W(\Phi)|$, given $\Phi'$ and $\Phi$. We use trace distance to quantify the error of two-qubit states, i.e., $D(\rho'_{AB}, \rho_{AB}):=\frac{1}{2}||\rho'_{AB}-\rho_{AB}|| \le \epsilon_{\rho}$, and diamond norm to quantify the error of a POVM, i.e., $D_{\Diamond}(\mathcal{M}',\mathcal{M}):=\max_{\tau}\frac{1}{2}||(\mathcal{M}'\otimes \mathbb{I})\tau-(\mathcal{M}\otimes \mathbb{I})\tau || \le \epsilon_{\mathcal{M}}$, where the maximization is done over all density matrices $\tau$ of dimension $2 \times 2$, and $\mathcal{M}$ specify the corresponding completely positive trace preserving (CPTP) map of a POVM. 

To calculate the upper bound of $\delta$, we focus on the difference of each element. Let $Q \in \{Q_{xy}\}$ and $Q=\Tr[\rho_{AB}A \otimes B]-\Tr[\rho_{A}A] \Tr[\rho_{B}B]$, where $A \in \{A_x\}$ and $B \in \{B_y\}$. Since any POVM could be viewed as a special map that maps a quantum state to a classical probability distribution, the classical distribution after acting $\mathcal{M}_{A \otimes B}$ on $\rho_{AB}$ is given by $\{p(a, b)\}$, where $\mathcal{M}_{A \otimes B}$ is the corresponding CPTP map of $A \otimes B$ (we will use similar notation for any POVM), $(a, b) \in \{(-1,-1),(1,-1),(-1,1),(1,1)\}$, and $p(-1, -1) +p(1, -1)+p(-1, 1)+ p(1, 1)=1$. After simple calculation, we have 
\begin{equation}
Q=4 [p(-1, -1) \times p(1, 1)-p(1, -1) \times p(-1, 1)].
\end{equation}
Similarly, we have the classical probability distribution $\{p'(a, b)\}$ after acting $\mathcal{M}_{A' \otimes B'}$ on $\rho'_{AB}$, where $A' \in \{A'_x\}$ and $B' \in \{B'_y\}$. The trace distance between $\{p(a, b)\}$ and $\{p'(a, b)\}$ is thus given by 
\begin{equation}
\label{D_p}
\begin{aligned}
D(p', p) := &\frac{1}{2}\sum_{a, b} |p'(a, b)-p(a, b)| \\
             =&D(\mathcal{M}_{A' \otimes B'} \rho'_{AB}, \mathcal{M}_{A \otimes B} \rho_{AB}) \\
             \le& D(\mathcal{M}_{A' \otimes B'} \rho'_{AB}, \mathcal{M}_{A \otimes B} \rho'_{AB})+D(\mathcal{M}_{A \otimes B} \rho'_{AB}, \mathcal{M}_{A \otimes B} \rho_{AB}) \\
             \le& D ( (\mathbb{I} \otimes \mathcal{M}_{B'}) (\mathcal{M}_{A'} \otimes \mathbb{I}) \rho'_{AB}, (\mathbb{I} \otimes \mathcal{M}_{B}) (\mathcal{M}_{A} \otimes \mathbb{I})  \rho'_{AB}) + \epsilon_{\rho} \\
             \le& D ( (\mathbb{I} \otimes \mathcal{M}_{B'}) (\mathcal{M}_{A'} \otimes \mathbb{I}) \rho'_{AB}, (\mathbb{I} \otimes \mathcal{M}_{B'}) (\mathcal{M}_{A} \otimes \mathbb{I}) \rho'_{AB}) \\
               &+D ( (\mathbb{I} \otimes \mathcal{M}_{B'}) (\mathcal{M}_{A} \otimes \mathbb{I}) \rho'_{AB}, (\mathbb{I} \otimes \mathcal{M}_{B}) (\mathcal{M}_{A} \otimes \mathbb{I}) \rho'_{AB}) + \epsilon_{\rho} \\
             \le& D ( (\mathcal{M}_{A'} \otimes \mathbb{I}) \rho'_{AB}, (\mathcal{M}_{A} \otimes \mathbb{I}) \rho'_{AB}) + \epsilon_{\mathcal{M}_{B}} + \epsilon_{\rho} \\
             \le& \epsilon_{\mathcal{M}_{A}}+\epsilon_{\mathcal{M}_{B}} + \epsilon_{\rho}, 
\end{aligned}
\end{equation}
where we use triangle inequality of trace distance in the third and fourth line, and the contraction under CPTP map in fourth, fifth and sixth line. Then, we could similarly define $Q'=\Tr[\rho'_{AB}A '\otimes B']-\Tr[\rho'_{A}A'] \Tr[\rho'_{B}B']$, where $\rho'_{A}=\Tr_B[\rho'_{AB}]$ and $\rho'_{B}=\Tr_A[\rho'_{AB}]$. Similarly,
\begin{equation}
Q'=4 [p'(-1, -1) \times p'(1, 1)-p'(1, -1) \times p'(-1, 1)].
\end{equation}
Then, we calculate the upper bound of $|Q'-Q|$. Obviously,
\begin{equation}
\label{p_difference}
\begin{aligned}
p'(-1, -1)p'(1, 1)-p(-1, -1)p(1, 1)&=[p'(-1, -1)-p(-1, -1)]p'(1, 1)+p(-1, -1)[p'(1, 1)-p(1, 1)] \\
                                                 &\le |p'(-1, -1)-p(-1, -1)| + |p'(1, 1)-p(1, 1)|,
\end{aligned}
\end{equation}
and similarly,
\begin{equation}
\begin{aligned}
p'(1, -1)p'(-1, 1)-p(1, -1)p(-1, 1) \le |p'(1, -1)-p(-1, 1)| + |p'(1, -1)-p(-1, 1)|,
\end{aligned}
\end{equation}
therefore, by Eq. \eqref{D_p}, we have
\begin{equation}
|Q'-Q| \le 8 D(p', p) \le 8(\epsilon_{\mathcal{M}_{A}}+\epsilon_{\mathcal{M}_{B}} + \epsilon_{\rho}).
\end{equation}

Now, we are ready to calculate the upper bound of $\delta$. Let $Q'_{xy}=\Tr[\rho'_{AB}A'_{x}\otimes B'_{y}]-\Tr[\rho'_{A}A'_x] \Tr[\rho'_{B}B'_y]$, then $W(\Phi')=Q'_{00}Q'_{11}-Q'_{01}Q'_{10}$. Thus, 
\begin{equation}
\begin{aligned}
\delta &=|W(\Phi')-W(\Phi)| \\
          &=|Q'_{00}Q'_{11}-Q_{00}Q_{11}-(Q'_{01}Q'_{10}-Q_{01}Q_{10})| \\
          &\le \sum_{x, y}|Q'_{xy}-Q_{xy}| \\
          &\le 16(\epsilon_{\mathcal{M}_{A_0}}+\epsilon_{\mathcal{M}_{B_0}} +\epsilon_{\mathcal{M}_{A_1}}+\epsilon_{\mathcal{M}_{B_1}} +2 \epsilon_{\rho}),
\end{aligned}
\end{equation}
where we make use of the same logic as Eq. \eqref{p_difference} in the third line. This result shows the continuity of our witness value. And the upper bound of $\delta$ grows linearly with parameters $\{\epsilon_{\mathcal{M}_{A_0}}, \epsilon_{\mathcal{M}_{B_0}}, \epsilon_{\mathcal{M}_{A_1}}, \epsilon_{\mathcal{M}_{B_1}}, \epsilon_{\rho} \}$, this shows the robustness of our method.


\section{Simulation}

In this section, we simulate the performance of our method. Here, we consider a model of channel loss, background error, finite and even basis-dependent detection efficiency in the two-qubit case. Let $\alpha_x$ be the total detection efficiency (including the channel efficiency and detector efficiency) on Alice's side, if the qubit vanishes before detection, or equivalently the detector outputs a "no click" or "double click" result, we label these outcomes as "-1" and "1" randomly, then we can equivalently view that the qubit goes through a depolarization channel featured by transmittance $\alpha_x$ before the POVM $A_x$. Besides, the term $a_x\mathbb{I}$ in $A_x$ actually characterize the background error since its resulting probability $\text{Tr}[a_x\mathbb{I} \rho]$ is irrelevant to the form of any qubit $\rho$. We combine the depolarization channel and the POVM $A_x$ together into another POVM $A^{'}_x$ given by 
\begin{equation}
A^{'}_x =a_x\mathbb{I}+\alpha_x \vec{A}_x \cdot \vec{\sigma},
\end{equation}
then we can see that $W \to \alpha_0 \alpha_1 W$ is strictly positive whenever $W>0$. We can have a same argument on Bob's side by considering the total detection efficiency $\beta_y$ and the term $b_y\mathbb{I}$.

\begin{figure}[htbp]
\centering
\includegraphics[width=0.5\linewidth]{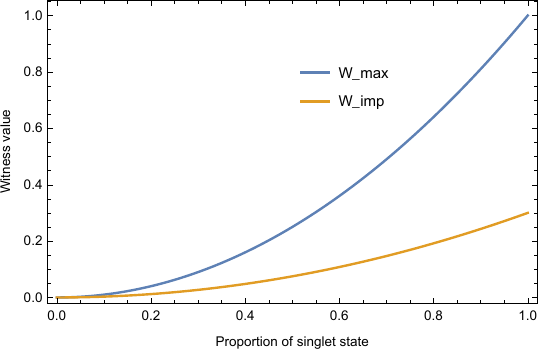}
\caption{For Werner state $p\ket{\phi^+}\bra{\phi^+}+\frac{1-p}{4}\mathbb{I}_4$: Maximize achievable witness $W_{\text{max}}$ and $W_{\text{imp}}$ when considering the imperfection of devices versus Proportion of singlet state $p$. Here, we set $\vec{A}_0=\vec{B}_0=\vec{x}$ and $\vec{A}_1=\vec{B}_1=\vec{z}$. For $W_{\text{max}}$, we set detection efficiency is $100\%$ on both sides to evolute it, for $W_{\text{imp}}$, we set Alice's detection efficiency is $\alpha_0=0.75$ and $\alpha_1=0.8$, Bob's detection efficiency is $\beta_0=0.8$ and $\beta_1=0.625$, to evolute it.}
\label{fig:simulation}
\end{figure}

Then, in Fig. \ref{fig:simulation}, we take the two-qubit Werner state as an example to simulate the behavior of maximal achievable witness $W_{\text{max}}$ and $W_{\text{imp}}$ when considering the imperfection of device. The Werner state is given by $p\ket{\phi^+}\bra{\phi^+}+\frac{1-p}{4}\mathbb{I}_4$, $\mathbb{I}_4$ is the identity matrix of four dimensional system and $p \in [0,1]$. The matrix $|\hat{S}|=|\hat{T}|$ with three eigenvalues $\{p, p, p\}$, then we can immediately have $W \le W_{\max}=p^2$. It is straightforward to see that an alternative way to achieving the maximal witness for the Werner state is to perform Pauli $\mathbb{X}$ and Pauli $\mathbb{Z}$ measurements on both Alice's and Bob's sides, that is, $\vec{A}_0=\vec{B}_0=\vec{x}$ and $\vec{A}_1=\vec{B}_1=\vec{z}$. Actually, for the Werner state, arbitrary two projective measurements as long as they are MUB could help us achieve the maximal witness, because $\mathbb{I}_4$ and $\ket{\phi^+}\bra{\phi^+}$ are invariant under local rotations. Here we also consider the case of imperfect devices, by setting $\vec{A}_0=\vec{B}_0=\vec{x}$ and $\vec{A}_1=\vec{B}_1=\vec{z}$, Alice's detection efficiency is $\alpha_0=0.75$ and $\alpha_1=0.8$, and Bob's detection efficiency is $\beta_0=0.8$ and $\beta_1=0.625$, we have $W_{\text{imp}}=\alpha_0 \alpha_1 \beta_0 \beta_1 W_{\text{max}}=0.3p^2$. In Fig. \ref{fig:simulation}, we can see that $W_{\max} >0 $ and $W_{\text{imp}} >0 $ always hold when discord is nonzero, that is $ p>0 $, which is different from the well-known entanglement condition of such state when $p>1/3$. Coincidentally, for the Werner state, $W_{\max}=p^2$ in our approach is same with the observable geometric discord \cite{dakic2010necessary} in Ref. \cite{girolami2012observable}, and differs with Ref. \cite{maziero2012classicality} only by a constant coefficient. Thus, while the previous method works, our method applies to uncharacterized measurements.


\section{Conclusion}

In this work, we propose a new approach to witnessing the QD for arbitrary dimension, and show that it is highly robust against the imperfections of devices. The high robustness and the fact that  $2 \times 2$ are sufficient for any bipartite state of arbitrarily large dimension shows our method is experimentally feasible.


\begin{acknowledgments}
We cordially thank Hoi-Kwong Lo for many helpful discussions.
R. W was supported by the University of Hong Kong start-up grant,
Y. Y is supported by National Natural Science Foundation of China (NSFC) (Grants No. 12175204),
Z. -Q. Y is supported by the National Natural Science Foundation of China (Grant Nos. 62171424, 61961136004).
\end{acknowledgments}

%
%
%
%
%
%
%
%
%
%
%
%
%
%
%

\nocite{*}

\bibliography{apssamp}

\providecommand{\noopsort}[1]{}\providecommand{\singleletter}[1]{#1}%
\begin{thebibliography}{36}%
\makeatletter
\providecommand \@ifxundefined [1]{%
 \@ifx{#1\undefined}
}%
\providecommand \@ifnum [1]{%
 \ifnum #1\expandafter \@firstoftwo
 \else \expandafter \@secondoftwo
 \fi
}%
\providecommand \@ifx [1]{%
 \ifx #1\expandafter \@firstoftwo
 \else \expandafter \@secondoftwo
 \fi
}%
\providecommand \natexlab [1]{#1}%
\providecommand \enquote  [1]{``#1''}%
\providecommand \bibnamefont  [1]{#1}%
\providecommand \bibfnamefont [1]{#1}%
\providecommand \citenamefont [1]{#1}%
\providecommand \href@noop [0]{\@secondoftwo}%
\providecommand \href [0]{\begingroup \@sanitize@url \@href}%
\providecommand \@href[1]{\@@startlink{#1}\@@href}%
\providecommand \@@href[1]{\endgroup#1\@@endlink}%
\providecommand \@sanitize@url [0]{\catcode `\\12\catcode `\$12\catcode
  `\&12\catcode `\#12\catcode `\^12\catcode `\_12\catcode `\%12\relax}%
\providecommand \@@startlink[1]{}%
\providecommand \@@endlink[0]{}%
\providecommand \url  [0]{\begingroup\@sanitize@url \@url }%
\providecommand \@url [1]{\endgroup\@href {#1}{\urlprefix }}%
\providecommand \urlprefix  [0]{URL }%
\providecommand \Eprint [0]{\href }%
\providecommand \doibase [0]{http://dx.doi.org/}%
\providecommand \selectlanguage [0]{\@gobble}%
\providecommand \bibinfo  [0]{\@secondoftwo}%
\providecommand \bibfield  [0]{\@secondoftwo}%
\providecommand \translation [1]{[#1]}%
\providecommand \BibitemOpen [0]{}%
\providecommand \bibitemStop [0]{}%
\providecommand \bibitemNoStop [0]{.\EOS\space}%
\providecommand \EOS [0]{\spacefactor3000\relax}%
\providecommand \BibitemShut  [1]{\csname bibitem#1\endcsname}%
\let\auto@bib@innerbib\@empty
\bibitem [{\citenamefont {Ollivier}\ and\ \citenamefont
  {Zurek}(2001)}]{ollivier2001quantum}%
  \BibitemOpen
  \bibfield  {author} {\bibinfo {author} {\bibfnamefont {H.}~\bibnamefont
  {Ollivier}}\ and\ \bibinfo {author} {\bibfnamefont {W.~H.}\ \bibnamefont
  {Zurek}},\ }\href@noop {} {\bibfield  {journal} {\bibinfo  {journal}
  {Physical review letters}\ }\textbf {\bibinfo {volume} {88}},\ \bibinfo
  {pages} {017901} (\bibinfo {year} {2001})}\BibitemShut {NoStop}%
\bibitem [{\citenamefont {Modi}\ \emph {et~al.}(2012)\citenamefont {Modi},
  \citenamefont {Brodutch}, \citenamefont {Cable}, \citenamefont {Paterek},\
  and\ \citenamefont {Vedral}}]{modi2012classical}%
  \BibitemOpen
  \bibfield  {author} {\bibinfo {author} {\bibfnamefont {K.}~\bibnamefont
  {Modi}}, \bibinfo {author} {\bibfnamefont {A.}~\bibnamefont {Brodutch}},
  \bibinfo {author} {\bibfnamefont {H.}~\bibnamefont {Cable}}, \bibinfo
  {author} {\bibfnamefont {T.}~\bibnamefont {Paterek}}, \ and\ \bibinfo
  {author} {\bibfnamefont {V.}~\bibnamefont {Vedral}},\ }\href@noop {}
  {\bibfield  {journal} {\bibinfo  {journal} {Reviews of Modern Physics}\
  }\textbf {\bibinfo {volume} {84}},\ \bibinfo {pages} {1655} (\bibinfo {year}
  {2012})}\BibitemShut {NoStop}%
\bibitem [{\citenamefont {Horodecki}\ \emph {et~al.}(2009)\citenamefont
  {Horodecki}, \citenamefont {Horodecki}, \citenamefont {Horodecki},\ and\
  \citenamefont {Horodecki}}]{horodecki2009quantum}%
  \BibitemOpen
  \bibfield  {author} {\bibinfo {author} {\bibfnamefont {R.}~\bibnamefont
  {Horodecki}}, \bibinfo {author} {\bibfnamefont {P.}~\bibnamefont
  {Horodecki}}, \bibinfo {author} {\bibfnamefont {M.}~\bibnamefont
  {Horodecki}}, \ and\ \bibinfo {author} {\bibfnamefont {K.}~\bibnamefont
  {Horodecki}},\ }\href@noop {} {\bibfield  {journal} {\bibinfo  {journal}
  {Reviews of modern physics}\ }\textbf {\bibinfo {volume} {81}},\ \bibinfo
  {pages} {865} (\bibinfo {year} {2009})}\BibitemShut {NoStop}%
\bibitem [{\citenamefont {Knill}\ and\ \citenamefont
  {Laflamme}(1998)}]{knill1998power}%
  \BibitemOpen
  \bibfield  {author} {\bibinfo {author} {\bibfnamefont {E.}~\bibnamefont
  {Knill}}\ and\ \bibinfo {author} {\bibfnamefont {R.}~\bibnamefont
  {Laflamme}},\ }\href@noop {} {\bibfield  {journal} {\bibinfo  {journal}
  {Physical Review Letters}\ }\textbf {\bibinfo {volume} {81}},\ \bibinfo
  {pages} {5672} (\bibinfo {year} {1998})}\BibitemShut {NoStop}%
\bibitem [{\citenamefont {Datta}\ \emph {et~al.}(2008)\citenamefont {Datta},
  \citenamefont {Shaji},\ and\ \citenamefont {Caves}}]{datta2008quantum}%
  \BibitemOpen
  \bibfield  {author} {\bibinfo {author} {\bibfnamefont {A.}~\bibnamefont
  {Datta}}, \bibinfo {author} {\bibfnamefont {A.}~\bibnamefont {Shaji}}, \ and\
  \bibinfo {author} {\bibfnamefont {C.~M.}\ \bibnamefont {Caves}},\ }\href@noop
  {} {\bibfield  {journal} {\bibinfo  {journal} {Physical review letters}\
  }\textbf {\bibinfo {volume} {100}},\ \bibinfo {pages} {050502} (\bibinfo
  {year} {2008})}\BibitemShut {NoStop}%
\bibitem [{\citenamefont {Daki{\'c}}\ \emph {et~al.}(2012)\citenamefont
  {Daki{\'c}}, \citenamefont {Lipp}, \citenamefont {Ma}, \citenamefont
  {Ringbauer}, \citenamefont {Kropatschek}, \citenamefont {Barz}, \citenamefont
  {Paterek}, \citenamefont {Vedral}, \citenamefont {Zeilinger}, \citenamefont
  {Brukner} \emph {et~al.}}]{dakic2012quantum}%
  \BibitemOpen
  \bibfield  {author} {\bibinfo {author} {\bibfnamefont {B.}~\bibnamefont
  {Daki{\'c}}}, \bibinfo {author} {\bibfnamefont {Y.~O.}\ \bibnamefont {Lipp}},
  \bibinfo {author} {\bibfnamefont {X.}~\bibnamefont {Ma}}, \bibinfo {author}
  {\bibfnamefont {M.}~\bibnamefont {Ringbauer}}, \bibinfo {author}
  {\bibfnamefont {S.}~\bibnamefont {Kropatschek}}, \bibinfo {author}
  {\bibfnamefont {S.}~\bibnamefont {Barz}}, \bibinfo {author} {\bibfnamefont
  {T.}~\bibnamefont {Paterek}}, \bibinfo {author} {\bibfnamefont
  {V.}~\bibnamefont {Vedral}}, \bibinfo {author} {\bibfnamefont
  {A.}~\bibnamefont {Zeilinger}}, \bibinfo {author} {\bibfnamefont
  {{\v{C}}.}~\bibnamefont {Brukner}},  \emph {et~al.},\ }\href@noop {}
  {\bibfield  {journal} {\bibinfo  {journal} {Nature Physics}\ }\textbf
  {\bibinfo {volume} {8}},\ \bibinfo {pages} {666} (\bibinfo {year}
  {2012})}\BibitemShut {NoStop}%
\bibitem [{\citenamefont {Gu}\ \emph {et~al.}(2012)\citenamefont {Gu},
  \citenamefont {Chrzanowski}, \citenamefont {Assad}, \citenamefont {Symul},
  \citenamefont {Modi}, \citenamefont {Ralph}, \citenamefont {Vedral},\ and\
  \citenamefont {Lam}}]{gu2012observing}%
  \BibitemOpen
  \bibfield  {author} {\bibinfo {author} {\bibfnamefont {M.}~\bibnamefont
  {Gu}}, \bibinfo {author} {\bibfnamefont {H.~M.}\ \bibnamefont {Chrzanowski}},
  \bibinfo {author} {\bibfnamefont {S.~M.}\ \bibnamefont {Assad}}, \bibinfo
  {author} {\bibfnamefont {T.}~\bibnamefont {Symul}}, \bibinfo {author}
  {\bibfnamefont {K.}~\bibnamefont {Modi}}, \bibinfo {author} {\bibfnamefont
  {T.~C.}\ \bibnamefont {Ralph}}, \bibinfo {author} {\bibfnamefont
  {V.}~\bibnamefont {Vedral}}, \ and\ \bibinfo {author} {\bibfnamefont {P.~K.}\
  \bibnamefont {Lam}},\ }\href@noop {} {\bibfield  {journal} {\bibinfo
  {journal} {Nature Physics}\ }\textbf {\bibinfo {volume} {8}},\ \bibinfo
  {pages} {671} (\bibinfo {year} {2012})}\BibitemShut {NoStop}%
\bibitem [{\citenamefont {Chuan}\ \emph {et~al.}(2012)\citenamefont {Chuan},
  \citenamefont {Maillard}, \citenamefont {Modi}, \citenamefont {Paterek},
  \citenamefont {Paternostro},\ and\ \citenamefont {Piani}}]{chuan2012quantum}%
  \BibitemOpen
  \bibfield  {author} {\bibinfo {author} {\bibfnamefont {T.}~\bibnamefont
  {Chuan}}, \bibinfo {author} {\bibfnamefont {J.}~\bibnamefont {Maillard}},
  \bibinfo {author} {\bibfnamefont {K.}~\bibnamefont {Modi}}, \bibinfo {author}
  {\bibfnamefont {T.}~\bibnamefont {Paterek}}, \bibinfo {author} {\bibfnamefont
  {M.}~\bibnamefont {Paternostro}}, \ and\ \bibinfo {author} {\bibfnamefont
  {M.}~\bibnamefont {Piani}},\ }\href@noop {} {\bibfield  {journal} {\bibinfo
  {journal} {Physical Review Letters}\ }\textbf {\bibinfo {volume} {109}},\
  \bibinfo {pages} {070501} (\bibinfo {year} {2012})}\BibitemShut {NoStop}%
\bibitem [{\citenamefont {Weedbrook}\ \emph {et~al.}(2016)\citenamefont
  {Weedbrook}, \citenamefont {Pirandola}, \citenamefont {Thompson},
  \citenamefont {Vedral},\ and\ \citenamefont {Gu}}]{weedbrook2016discord}%
  \BibitemOpen
  \bibfield  {author} {\bibinfo {author} {\bibfnamefont {C.}~\bibnamefont
  {Weedbrook}}, \bibinfo {author} {\bibfnamefont {S.}~\bibnamefont
  {Pirandola}}, \bibinfo {author} {\bibfnamefont {J.}~\bibnamefont {Thompson}},
  \bibinfo {author} {\bibfnamefont {V.}~\bibnamefont {Vedral}}, \ and\ \bibinfo
  {author} {\bibfnamefont {M.}~\bibnamefont {Gu}},\ }\href@noop {} {\bibfield
  {journal} {\bibinfo  {journal} {New Journal of Physics}\ }\textbf {\bibinfo
  {volume} {18}},\ \bibinfo {pages} {043027} (\bibinfo {year}
  {2016})}\BibitemShut {NoStop}%
\bibitem [{\citenamefont {Girolami}\ \emph {et~al.}(2013)\citenamefont
  {Girolami}, \citenamefont {Tufarelli},\ and\ \citenamefont
  {Adesso}}]{girolami2013characterizing}%
  \BibitemOpen
  \bibfield  {author} {\bibinfo {author} {\bibfnamefont {D.}~\bibnamefont
  {Girolami}}, \bibinfo {author} {\bibfnamefont {T.}~\bibnamefont {Tufarelli}},
  \ and\ \bibinfo {author} {\bibfnamefont {G.}~\bibnamefont {Adesso}},\
  }\href@noop {} {\bibfield  {journal} {\bibinfo  {journal} {Physical review
  letters}\ }\textbf {\bibinfo {volume} {110}},\ \bibinfo {pages} {240402}
  (\bibinfo {year} {2013})}\BibitemShut {NoStop}%
\bibitem [{\citenamefont {Cavalcanti}\ \emph {et~al.}(2011)\citenamefont
  {Cavalcanti}, \citenamefont {Aolita}, \citenamefont {Boixo}, \citenamefont
  {Modi}, \citenamefont {Piani},\ and\ \citenamefont
  {Winter}}]{cavalcanti2011operational}%
  \BibitemOpen
  \bibfield  {author} {\bibinfo {author} {\bibfnamefont {D.}~\bibnamefont
  {Cavalcanti}}, \bibinfo {author} {\bibfnamefont {L.}~\bibnamefont {Aolita}},
  \bibinfo {author} {\bibfnamefont {S.}~\bibnamefont {Boixo}}, \bibinfo
  {author} {\bibfnamefont {K.}~\bibnamefont {Modi}}, \bibinfo {author}
  {\bibfnamefont {M.}~\bibnamefont {Piani}}, \ and\ \bibinfo {author}
  {\bibfnamefont {A.}~\bibnamefont {Winter}},\ }\href@noop {} {\bibfield
  {journal} {\bibinfo  {journal} {Physical Review A}\ }\textbf {\bibinfo
  {volume} {83}},\ \bibinfo {pages} {032324} (\bibinfo {year}
  {2011})}\BibitemShut {NoStop}%
\bibitem [{\citenamefont {Madhok}\ and\ \citenamefont
  {Datta}(2011)}]{madhok2011interpreting}%
  \BibitemOpen
  \bibfield  {author} {\bibinfo {author} {\bibfnamefont {V.}~\bibnamefont
  {Madhok}}\ and\ \bibinfo {author} {\bibfnamefont {A.}~\bibnamefont {Datta}},\
  }\href@noop {} {\bibfield  {journal} {\bibinfo  {journal} {Physical Review
  A}\ }\textbf {\bibinfo {volume} {83}},\ \bibinfo {pages} {032323} (\bibinfo
  {year} {2011})}\BibitemShut {NoStop}%
\bibitem [{\citenamefont {Yu}\ and\ \citenamefont
  {Eberly}(2009)}]{yu2009sudden}%
  \BibitemOpen
  \bibfield  {author} {\bibinfo {author} {\bibfnamefont {T.}~\bibnamefont
  {Yu}}\ and\ \bibinfo {author} {\bibfnamefont {J.}~\bibnamefont {Eberly}},\
  }\href@noop {} {\bibfield  {journal} {\bibinfo  {journal} {Science}\ }\textbf
  {\bibinfo {volume} {323}},\ \bibinfo {pages} {598} (\bibinfo {year}
  {2009})}\BibitemShut {NoStop}%
\bibitem [{\citenamefont {Ferraro}\ \emph {et~al.}(2010)\citenamefont
  {Ferraro}, \citenamefont {Aolita}, \citenamefont {Cavalcanti}, \citenamefont
  {Cucchietti},\ and\ \citenamefont {Ac{\'\i}n}}]{ferraro2010almost}%
  \BibitemOpen
  \bibfield  {author} {\bibinfo {author} {\bibfnamefont {A.}~\bibnamefont
  {Ferraro}}, \bibinfo {author} {\bibfnamefont {L.}~\bibnamefont {Aolita}},
  \bibinfo {author} {\bibfnamefont {D.}~\bibnamefont {Cavalcanti}}, \bibinfo
  {author} {\bibfnamefont {F.~M.}\ \bibnamefont {Cucchietti}}, \ and\ \bibinfo
  {author} {\bibfnamefont {A.}~\bibnamefont {Ac{\'\i}n}},\ }\href@noop {}
  {\bibfield  {journal} {\bibinfo  {journal} {Physical Review A}\ }\textbf
  {\bibinfo {volume} {81}},\ \bibinfo {pages} {052318} (\bibinfo {year}
  {2010})}\BibitemShut {NoStop}%
\bibitem [{\citenamefont {Chitambar}\ and\ \citenamefont
  {Gour}(2019)}]{chitambar2019quantum}%
  \BibitemOpen
  \bibfield  {author} {\bibinfo {author} {\bibfnamefont {E.}~\bibnamefont
  {Chitambar}}\ and\ \bibinfo {author} {\bibfnamefont {G.}~\bibnamefont
  {Gour}},\ }\href@noop {} {\bibfield  {journal} {\bibinfo  {journal} {Reviews
  of Modern Physics}\ }\textbf {\bibinfo {volume} {91}},\ \bibinfo {pages}
  {025001} (\bibinfo {year} {2019})}\BibitemShut {NoStop}%
\bibitem [{\citenamefont {Bera}\ \emph {et~al.}(2017)\citenamefont {Bera},
  \citenamefont {Das}, \citenamefont {Sadhukhan}, \citenamefont {Roy},
  \citenamefont {De},\ and\ \citenamefont {Sen}}]{bera2017quantum}%
  \BibitemOpen
  \bibfield  {author} {\bibinfo {author} {\bibfnamefont {A.}~\bibnamefont
  {Bera}}, \bibinfo {author} {\bibfnamefont {T.}~\bibnamefont {Das}}, \bibinfo
  {author} {\bibfnamefont {D.}~\bibnamefont {Sadhukhan}}, \bibinfo {author}
  {\bibfnamefont {S.~S.}\ \bibnamefont {Roy}}, \bibinfo {author} {\bibfnamefont
  {A.~S.}\ \bibnamefont {De}}, \ and\ \bibinfo {author} {\bibfnamefont
  {U.}~\bibnamefont {Sen}},\ }\href@noop {} {\bibfield  {journal} {\bibinfo
  {journal} {Reports on Progress in Physics}\ }\textbf {\bibinfo {volume}
  {81}},\ \bibinfo {pages} {024001} (\bibinfo {year} {2017})}\BibitemShut
  {NoStop}%
\bibitem [{\citenamefont {Rahimi}\ and\ \citenamefont
  {SaiToh}(2010)}]{rahimi2010single}%
  \BibitemOpen
  \bibfield  {author} {\bibinfo {author} {\bibfnamefont {R.}~\bibnamefont
  {Rahimi}}\ and\ \bibinfo {author} {\bibfnamefont {A.}~\bibnamefont
  {SaiToh}},\ }\href@noop {} {\bibfield  {journal} {\bibinfo  {journal}
  {Physical Review A}\ }\textbf {\bibinfo {volume} {82}},\ \bibinfo {pages}
  {022314} (\bibinfo {year} {2010})}\BibitemShut {NoStop}%
\bibitem [{\citenamefont {Bylicka}\ and\ \citenamefont
  {Chru{\'s}ci{\'n}ski}(2010)}]{bylicka2010witnessing}%
  \BibitemOpen
  \bibfield  {author} {\bibinfo {author} {\bibfnamefont {B.}~\bibnamefont
  {Bylicka}}\ and\ \bibinfo {author} {\bibfnamefont {D.}~\bibnamefont
  {Chru{\'s}ci{\'n}ski}},\ }\href@noop {} {\bibfield  {journal} {\bibinfo
  {journal} {Physical Review A}\ }\textbf {\bibinfo {volume} {81}},\ \bibinfo
  {pages} {062102} (\bibinfo {year} {2010})}\BibitemShut {NoStop}%
\bibitem [{\citenamefont {Gessner}\ and\ \citenamefont
  {Breuer}(2013)}]{gessner2013local}%
  \BibitemOpen
  \bibfield  {author} {\bibinfo {author} {\bibfnamefont {M.}~\bibnamefont
  {Gessner}}\ and\ \bibinfo {author} {\bibfnamefont {H.-P.}\ \bibnamefont
  {Breuer}},\ }\href@noop {} {\bibfield  {journal} {\bibinfo  {journal}
  {Physical Review A}\ }\textbf {\bibinfo {volume} {87}},\ \bibinfo {pages}
  {042107} (\bibinfo {year} {2013})}\BibitemShut {NoStop}%
\bibitem [{\citenamefont {Zhang}\ \emph {et~al.}(2011)\citenamefont {Zhang},
  \citenamefont {Yu}, \citenamefont {Chen},\ and\ \citenamefont
  {Oh}}]{zhang2011detecting}%
  \BibitemOpen
  \bibfield  {author} {\bibinfo {author} {\bibfnamefont {C.}~\bibnamefont
  {Zhang}}, \bibinfo {author} {\bibfnamefont {S.}~\bibnamefont {Yu}}, \bibinfo
  {author} {\bibfnamefont {Q.}~\bibnamefont {Chen}}, \ and\ \bibinfo {author}
  {\bibfnamefont {C.}~\bibnamefont {Oh}},\ }\href@noop {} {\bibfield  {journal}
  {\bibinfo  {journal} {Physical Review A}\ }\textbf {\bibinfo {volume} {84}},\
  \bibinfo {pages} {032122} (\bibinfo {year} {2011})}\BibitemShut {NoStop}%
\bibitem [{\citenamefont {Maziero}\ and\ \citenamefont
  {Serra}(2012)}]{maziero2012classicality}%
  \BibitemOpen
  \bibfield  {author} {\bibinfo {author} {\bibfnamefont {J.}~\bibnamefont
  {Maziero}}\ and\ \bibinfo {author} {\bibfnamefont {R.~M.}\ \bibnamefont
  {Serra}},\ }\href@noop {} {\bibfield  {journal} {\bibinfo  {journal}
  {International Journal of Quantum Information}\ }\textbf {\bibinfo {volume}
  {10}},\ \bibinfo {pages} {1250028} (\bibinfo {year} {2012})}\BibitemShut
  {NoStop}%
\bibitem [{\citenamefont {Girolami}\ and\ \citenamefont
  {Adesso}(2012)}]{girolami2012observable}%
  \BibitemOpen
  \bibfield  {author} {\bibinfo {author} {\bibfnamefont {D.}~\bibnamefont
  {Girolami}}\ and\ \bibinfo {author} {\bibfnamefont {G.}~\bibnamefont
  {Adesso}},\ }\href@noop {} {\bibfield  {journal} {\bibinfo  {journal}
  {Physical review letters}\ }\textbf {\bibinfo {volume} {108}},\ \bibinfo
  {pages} {150403} (\bibinfo {year} {2012})}\BibitemShut {NoStop}%
\bibitem [{\citenamefont {Xu}\ \emph {et~al.}(2010)\citenamefont {Xu},
  \citenamefont {Xu}, \citenamefont {Li}, \citenamefont {Zhang}, \citenamefont
  {Zou},\ and\ \citenamefont {Guo}}]{xu2010experimental}%
  \BibitemOpen
  \bibfield  {author} {\bibinfo {author} {\bibfnamefont {J.-S.}\ \bibnamefont
  {Xu}}, \bibinfo {author} {\bibfnamefont {X.-Y.}\ \bibnamefont {Xu}}, \bibinfo
  {author} {\bibfnamefont {C.-F.}\ \bibnamefont {Li}}, \bibinfo {author}
  {\bibfnamefont {C.-J.}\ \bibnamefont {Zhang}}, \bibinfo {author}
  {\bibfnamefont {X.-B.}\ \bibnamefont {Zou}}, \ and\ \bibinfo {author}
  {\bibfnamefont {G.-C.}\ \bibnamefont {Guo}},\ }\href@noop {} {\bibfield
  {journal} {\bibinfo  {journal} {Nature communications}\ }\textbf {\bibinfo
  {volume} {1}},\ \bibinfo {pages} {1} (\bibinfo {year} {2010})}\BibitemShut
  {NoStop}%
\bibitem [{\citenamefont {Passante}\ \emph {et~al.}(2011)\citenamefont
  {Passante}, \citenamefont {Moussa}, \citenamefont {Trottier},\ and\
  \citenamefont {Laflamme}}]{passante2011experimental}%
  \BibitemOpen
  \bibfield  {author} {\bibinfo {author} {\bibfnamefont {G.}~\bibnamefont
  {Passante}}, \bibinfo {author} {\bibfnamefont {O.}~\bibnamefont {Moussa}},
  \bibinfo {author} {\bibfnamefont {D.}~\bibnamefont {Trottier}}, \ and\
  \bibinfo {author} {\bibfnamefont {R.}~\bibnamefont {Laflamme}},\ }\href@noop
  {} {\bibfield  {journal} {\bibinfo  {journal} {Physical Review A}\ }\textbf
  {\bibinfo {volume} {84}},\ \bibinfo {pages} {044302} (\bibinfo {year}
  {2011})}\BibitemShut {NoStop}%
\bibitem [{\citenamefont {Aguilar}\ \emph {et~al.}(2012)\citenamefont
  {Aguilar}, \citenamefont {Far{\'\i}as}, \citenamefont {Maziero},
  \citenamefont {Serra}, \citenamefont {Ribeiro},\ and\ \citenamefont
  {Walborn}}]{aguilar2012experimental}%
  \BibitemOpen
  \bibfield  {author} {\bibinfo {author} {\bibfnamefont {G.}~\bibnamefont
  {Aguilar}}, \bibinfo {author} {\bibfnamefont {O.~J.}\ \bibnamefont
  {Far{\'\i}as}}, \bibinfo {author} {\bibfnamefont {J.}~\bibnamefont
  {Maziero}}, \bibinfo {author} {\bibfnamefont {R.}~\bibnamefont {Serra}},
  \bibinfo {author} {\bibfnamefont {P.~S.}\ \bibnamefont {Ribeiro}}, \ and\
  \bibinfo {author} {\bibfnamefont {S.}~\bibnamefont {Walborn}},\ }\href@noop
  {} {\bibfield  {journal} {\bibinfo  {journal} {Physical Review Letters}\
  }\textbf {\bibinfo {volume} {108}},\ \bibinfo {pages} {063601} (\bibinfo
  {year} {2012})}\BibitemShut {NoStop}%
\bibitem [{\citenamefont {Silva}\ \emph {et~al.}(2013)\citenamefont {Silva},
  \citenamefont {Girolami}, \citenamefont {Auccaise}, \citenamefont {Sarthour},
  \citenamefont {Oliveira}, \citenamefont {Bonagamba}, \citenamefont
  {Deazevedo}, \citenamefont {Soares-Pinto},\ and\ \citenamefont
  {Adesso}}]{silva2013measuring}%
  \BibitemOpen
  \bibfield  {author} {\bibinfo {author} {\bibfnamefont {I.}~\bibnamefont
  {Silva}}, \bibinfo {author} {\bibfnamefont {D.}~\bibnamefont {Girolami}},
  \bibinfo {author} {\bibfnamefont {R.}~\bibnamefont {Auccaise}}, \bibinfo
  {author} {\bibfnamefont {R.}~\bibnamefont {Sarthour}}, \bibinfo {author}
  {\bibfnamefont {I.}~\bibnamefont {Oliveira}}, \bibinfo {author}
  {\bibfnamefont {T.~J.}\ \bibnamefont {Bonagamba}}, \bibinfo {author}
  {\bibfnamefont {E.}~\bibnamefont {Deazevedo}}, \bibinfo {author}
  {\bibfnamefont {D.}~\bibnamefont {Soares-Pinto}}, \ and\ \bibinfo {author}
  {\bibfnamefont {G.}~\bibnamefont {Adesso}},\ }\href@noop {} {\bibfield
  {journal} {\bibinfo  {journal} {Physical Review Letters}\ }\textbf {\bibinfo
  {volume} {110}},\ \bibinfo {pages} {140501} (\bibinfo {year}
  {2013})}\BibitemShut {NoStop}%
\bibitem [{\citenamefont {Bowles}\ \emph {et~al.}(2014)\citenamefont {Bowles},
  \citenamefont {Quintino},\ and\ \citenamefont
  {Brunner}}]{bowles2014certifying}%
  \BibitemOpen
  \bibfield  {author} {\bibinfo {author} {\bibfnamefont {J.}~\bibnamefont
  {Bowles}}, \bibinfo {author} {\bibfnamefont {M.~T.}\ \bibnamefont
  {Quintino}}, \ and\ \bibinfo {author} {\bibfnamefont {N.}~\bibnamefont
  {Brunner}},\ }\href@noop {} {\bibfield  {journal} {\bibinfo  {journal}
  {Physical review letters}\ }\textbf {\bibinfo {volume} {112}},\ \bibinfo
  {pages} {140407} (\bibinfo {year} {2014})}\BibitemShut {NoStop}%
\bibitem [{\citenamefont {Datta}(2008)}]{datta2008studies}%
  \BibitemOpen
  \bibfield  {author} {\bibinfo {author} {\bibfnamefont {A.}~\bibnamefont
  {Datta}},\ }\href@noop {} {\emph {\bibinfo {title} {Studies on the role of
  entanglement in mixed-state quantum computation}}}\ (\bibinfo  {publisher}
  {The University of New Mexico},\ \bibinfo {year} {2008})\BibitemShut
  {NoStop}%
\bibitem [{\citenamefont {Dittmer}(1994)}]{dittmer1994cross}%
  \BibitemOpen
  \bibfield  {author} {\bibinfo {author} {\bibfnamefont {A.}~\bibnamefont
  {Dittmer}},\ }\href@noop {} {\bibfield  {journal} {\bibinfo  {journal} {The
  American Mathematical Monthly}\ }\textbf {\bibinfo {volume} {101}},\ \bibinfo
  {pages} {887} (\bibinfo {year} {1994})}\BibitemShut {NoStop}%
\bibitem [{\citenamefont {Erhard}\ \emph {et~al.}(2020)\citenamefont {Erhard},
  \citenamefont {Krenn},\ and\ \citenamefont {Zeilinger}}]{erhard2020advances}%
  \BibitemOpen
  \bibfield  {author} {\bibinfo {author} {\bibfnamefont {M.}~\bibnamefont
  {Erhard}}, \bibinfo {author} {\bibfnamefont {M.}~\bibnamefont {Krenn}}, \
  and\ \bibinfo {author} {\bibfnamefont {A.}~\bibnamefont {Zeilinger}},\
  }\href@noop {} {\bibfield  {journal} {\bibinfo  {journal} {Nature Reviews
  Physics}\ }\textbf {\bibinfo {volume} {2}},\ \bibinfo {pages} {365} (\bibinfo
  {year} {2020})}\BibitemShut {NoStop}%
\bibitem [{\citenamefont {Buscemi}\ \emph {et~al.}(2005)\citenamefont
  {Buscemi}, \citenamefont {Keyl}, \citenamefont {D’Ariano}, \citenamefont
  {Perinotti},\ and\ \citenamefont {Werner}}]{buscemi2005clean}%
  \BibitemOpen
  \bibfield  {author} {\bibinfo {author} {\bibfnamefont {F.}~\bibnamefont
  {Buscemi}}, \bibinfo {author} {\bibfnamefont {M.}~\bibnamefont {Keyl}},
  \bibinfo {author} {\bibfnamefont {G.~M.}\ \bibnamefont {D’Ariano}},
  \bibinfo {author} {\bibfnamefont {P.}~\bibnamefont {Perinotti}}, \ and\
  \bibinfo {author} {\bibfnamefont {R.~F.}\ \bibnamefont {Werner}},\
  }\href@noop {} {\bibfield  {journal} {\bibinfo  {journal} {Journal of
  mathematical physics}\ }\textbf {\bibinfo {volume} {46}},\ \bibinfo {pages}
  {082109} (\bibinfo {year} {2005})}\BibitemShut {NoStop}%
\bibitem [{\citenamefont {D'Ariano}\ \emph {et~al.}(2005)\citenamefont
  {D'Ariano}, \citenamefont {Presti},\ and\ \citenamefont
  {Perinotti}}]{d2005classical}%
  \BibitemOpen
  \bibfield  {author} {\bibinfo {author} {\bibfnamefont {G.~M.}\ \bibnamefont
  {D'Ariano}}, \bibinfo {author} {\bibfnamefont {P.~L.}\ \bibnamefont
  {Presti}}, \ and\ \bibinfo {author} {\bibfnamefont {P.}~\bibnamefont
  {Perinotti}},\ }\href@noop {} {\bibfield  {journal} {\bibinfo  {journal}
  {Journal of Physics A: Mathematical and General}\ }\textbf {\bibinfo {volume}
  {38}},\ \bibinfo {pages} {5979} (\bibinfo {year} {2005})}\BibitemShut
  {NoStop}%
\bibitem [{\citenamefont {Haapasalo}\ \emph {et~al.}(2012)\citenamefont
  {Haapasalo}, \citenamefont {Heinosaari},\ and\ \citenamefont
  {Pellonp{\"a}{\"a}}}]{haapasalo2012quantum}%
  \BibitemOpen
  \bibfield  {author} {\bibinfo {author} {\bibfnamefont {E.}~\bibnamefont
  {Haapasalo}}, \bibinfo {author} {\bibfnamefont {T.}~\bibnamefont
  {Heinosaari}}, \ and\ \bibinfo {author} {\bibfnamefont {J.-P.}\ \bibnamefont
  {Pellonp{\"a}{\"a}}},\ }\href@noop {} {\bibfield  {journal} {\bibinfo
  {journal} {Quantum Information Processing}\ }\textbf {\bibinfo {volume}
  {11}},\ \bibinfo {pages} {1751} (\bibinfo {year} {2012})}\BibitemShut
  {NoStop}%
\bibitem [{\citenamefont {Oszmaniec}\ \emph {et~al.}(2017)\citenamefont
  {Oszmaniec}, \citenamefont {Guerini}, \citenamefont {Wittek},\ and\
  \citenamefont {Ac{\'\i}n}}]{oszmaniec2017simulating}%
  \BibitemOpen
  \bibfield  {author} {\bibinfo {author} {\bibfnamefont {M.}~\bibnamefont
  {Oszmaniec}}, \bibinfo {author} {\bibfnamefont {L.}~\bibnamefont {Guerini}},
  \bibinfo {author} {\bibfnamefont {P.}~\bibnamefont {Wittek}}, \ and\ \bibinfo
  {author} {\bibfnamefont {A.}~\bibnamefont {Ac{\'\i}n}},\ }\href@noop {}
  {\bibfield  {journal} {\bibinfo  {journal} {Physical review letters}\
  }\textbf {\bibinfo {volume} {119}},\ \bibinfo {pages} {190501} (\bibinfo
  {year} {2017})}\BibitemShut {NoStop}%
\bibitem [{\citenamefont {Daki{\'c}}\ \emph {et~al.}(2010)\citenamefont
  {Daki{\'c}}, \citenamefont {Vedral},\ and\ \citenamefont
  {Brukner}}]{dakic2010necessary}%
  \BibitemOpen
  \bibfield  {author} {\bibinfo {author} {\bibfnamefont {B.}~\bibnamefont
  {Daki{\'c}}}, \bibinfo {author} {\bibfnamefont {V.}~\bibnamefont {Vedral}}, \
  and\ \bibinfo {author} {\bibfnamefont {{\v{C}}.}~\bibnamefont {Brukner}},\
  }\href@noop {} {\bibfield  {journal} {\bibinfo  {journal} {Physical review
  letters}\ }\textbf {\bibinfo {volume} {105}},\ \bibinfo {pages} {190502}
  (\bibinfo {year} {2010})}\BibitemShut {NoStop}%
\bibitem [{\citenamefont {Fanchini}\ \emph {et~al.}(2011)\citenamefont
  {Fanchini}, \citenamefont {Cornelio}, \citenamefont {de~Oliveira},\ and\
  \citenamefont {Caldeira}}]{fanchini2011conservation}%
  \BibitemOpen
  \bibfield  {author} {\bibinfo {author} {\bibfnamefont {F.~F.}\ \bibnamefont
  {Fanchini}}, \bibinfo {author} {\bibfnamefont {M.~F.}\ \bibnamefont
  {Cornelio}}, \bibinfo {author} {\bibfnamefont {M.~C.}\ \bibnamefont
  {de~Oliveira}}, \ and\ \bibinfo {author} {\bibfnamefont {A.~O.}\ \bibnamefont
  {Caldeira}},\ }\href@noop {} {\bibfield  {journal} {\bibinfo  {journal}
  {Physical Review A}\ }\textbf {\bibinfo {volume} {84}},\ \bibinfo {pages}
  {012313} (\bibinfo {year} {2011})}\BibitemShut {NoStop}%
\end{thebibliography}%

\end{document}